\documentclass[a4paper,fleqn,usenatbib]{mnras}
\usepackage[T1]{fontenc}
\usepackage{ae,aecompl}

\usepackage{graphicx}	% Including figure files
\usepackage{amsmath}	% Advanced maths commands
\usepackage{amssymb}	% Extra maths symbols

\usepackage{color}
\usepackage{journals}

\title[Interplay of spectral components]
      {Interplay of spectral components in timing properties of accreting compact objects}
 \author[Alexandra~Veledina]
{Alexandra~Veledina\thanks{E-mail: alexandra.veledina@gmail.com} \\ 
Department of Physics and Astronomy, FI-20014 University of Turku, Finland\\
Nordita, KTH Royal Institute of Technology and Stockholm University, Roslagstullsbacken 23, SE-10691 Stockholm, Sweden\\
Space Research Institute of the Russian Academy of Sciences, Profsoyuznaya Str. 84/32, Moscow 117997, Russia\\}

\begin{document}

\maketitle

\begin{abstract}
\noindent
X-ray variability of accreting black hole binaries is believed to be produced by the fluctuations propagating towards the compact object.
Observations suggest the light curves in different energy bands are connected, but the fluctuations at harder energies are delayed with respect to the softer ones.
The standard interpretation involves dependence of the X-ray spectral hardness on the radial distance, with harder spectra emitted closer to the black hole. 
Recently, a number of challenges to this scenario have been found, both at qualitative and quantitative level.
The model does not predict the large magnitude of the delay between different energies, as well as the dependence of variability amplitude on the spectral range and frequency.
We study timing properties of accreting black hole X-ray binaries taking into account peculiarities of spectral formation in these sources.
We show that the simultaneous presence of two components leads to a complex shape of power spectra and phase lags between the X-ray bands:
the amplitude of the phase lag becomes independent of the delay between the components, and the power spectra show an artificial decrease or, on the contrary, enhancement of power at low Fourier frequencies.
This provides an essential ingredient for the reinstatement of the propagating fluctuations model. 
The work draws attention to the importance of considering conditions of spectral formation when studying timing properties.
\end{abstract}

\begin{keywords}
{accretion, accretion discs -- X-rays: binaries }
 \end{keywords}

\section{Introduction}

The majority of accreting black hole binaries are transients. 
They undergo outbursts lasting weeks to months, and then reside in quiescence for years to decades. 
During the outbursts the objects demonstrate evolution of the X-ray spectra from hard (when maximum energy is emitted in the hard X-rays) to soft (with the peak energy around 1~keV) and then back to the hard state, in addition to the changing X-ray luminosity \citep{RM06}.
The transitions between the hard and soft states proceeds through the so-called hard-intermediate and soft-intermediate states, identified most easily by the timing properties \citep{BM16}.

There is an overall consensus on geometry and spectral formation in the soft state: emission above $\sim$3~keV is mostly produced by Compton up-scattering off the electrons in the hot corona  \citep{PC98,Gier99} atop of the standard cold accretion disc  \citep{SS73,NT73}.
The Comptonization is fed in this case by the soft X-ray emission of the underlying accretion disc.
Less is certain about the hard state, and there have been debates on whether the cold disc goes all the way down to the innermost stable circular orbit or is truncated at some radius \citep{ZG04,DGK07}.
The related question concerns the source of seed photons in this state, whether those are provided by the cold disc \citep{PKR97} or by the synchrotron emission from the hot accretion flow itself \citep{Esin97,PV14}.
The scenario where the disc is providing the seed photons suffers substantial difficulties in reproducing the stable spectral shape and electron temperature, as well as in terms of multiwavelength behaviour \citep{PVZ18}. 
On the other hand, all of them naturally appear in the scenario with synchrotron Comptonization \citep{VP09,MB09,VPV13,PV14}.

If the synchrotron mechanism is the major source of seed photons in the hard state and the cold disc is the dominant source in the soft state, there has to be a transitional region where they switch.
If the switch occurs, it is natural to expect it to happen during the intermediate states.
Interestingly, spectral modelling suggests it is sometimes not possible to fit the data using one Comptonization continuum, and multiple (at least, two) continua are required \citep{IPG05,SUT11,Yamada13,AD18}.
These two continua can be produced by the two sources of seed photons, one coming from the accretion disc, and the other coming from synchrotron radiation.

Recently, a critical luminosity at which the two sources switch was identified in the source SWIFT~J1753.5--0127 \citep{KVT16}. 
As the source declines towards the quiescence, both Comptonization and disc luminosities decrease, and the spectral hardness has to evolve according to the amount of seed photons.
Yet, the substantial decrease of the disc luminosity was not followed by the corresponding hardening of the spectrum in the source, as if the Comptonization continuum contains much more seed photons than the disc itself. 
For the photon-starved Comptonization this situation is impossible, hence there has to exist an additional source of seed photons. 
Similar arguments, but from the point of view of accretion rates through the disc and through the corona, came to the same conclusion \citep{CDSG10}.

% (Fig. 1)
\begin{figure*}
\centering 
\includegraphics[width=7cm]{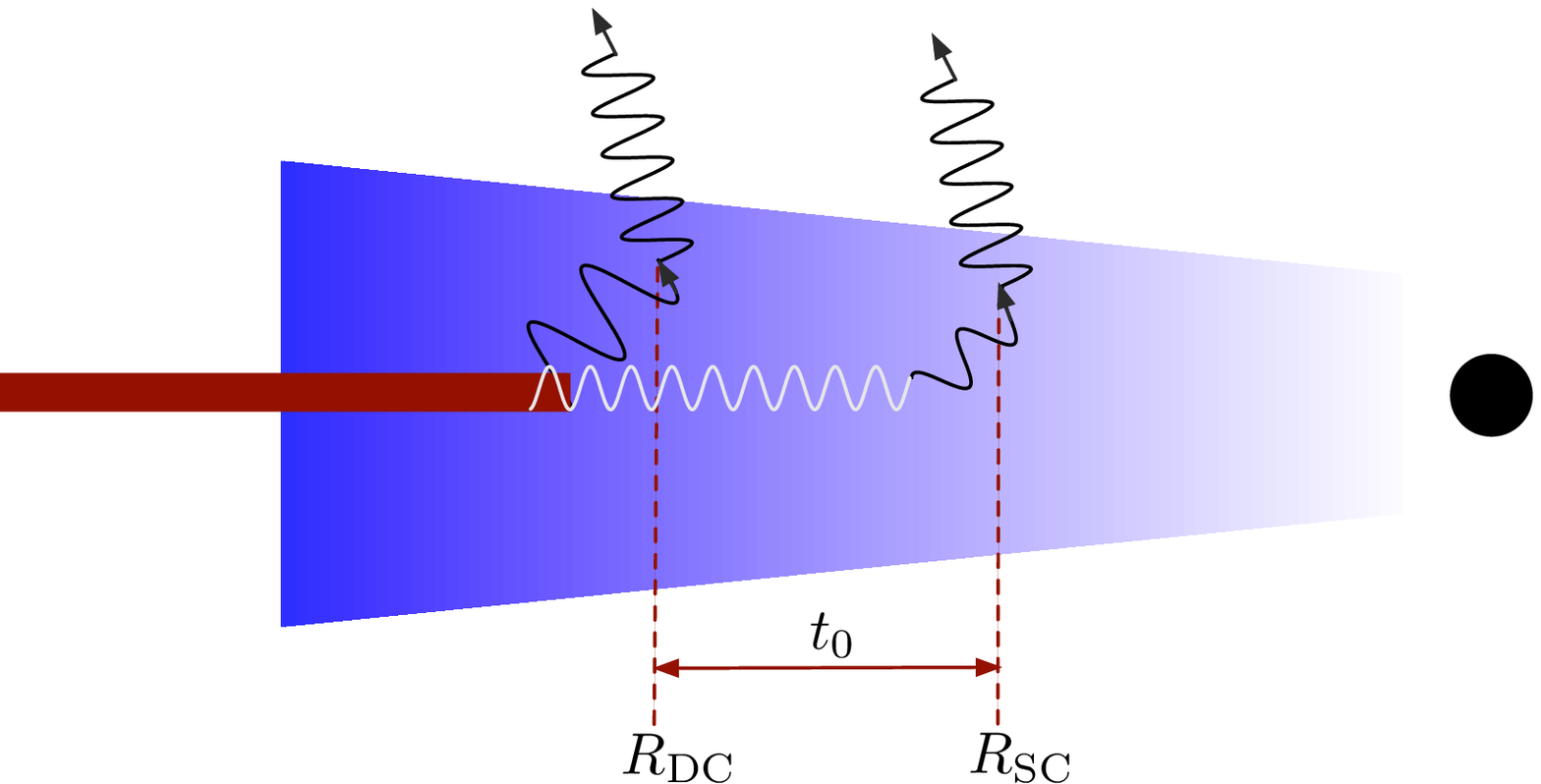}
\hspace{1 cm}
\includegraphics[width=7cm]{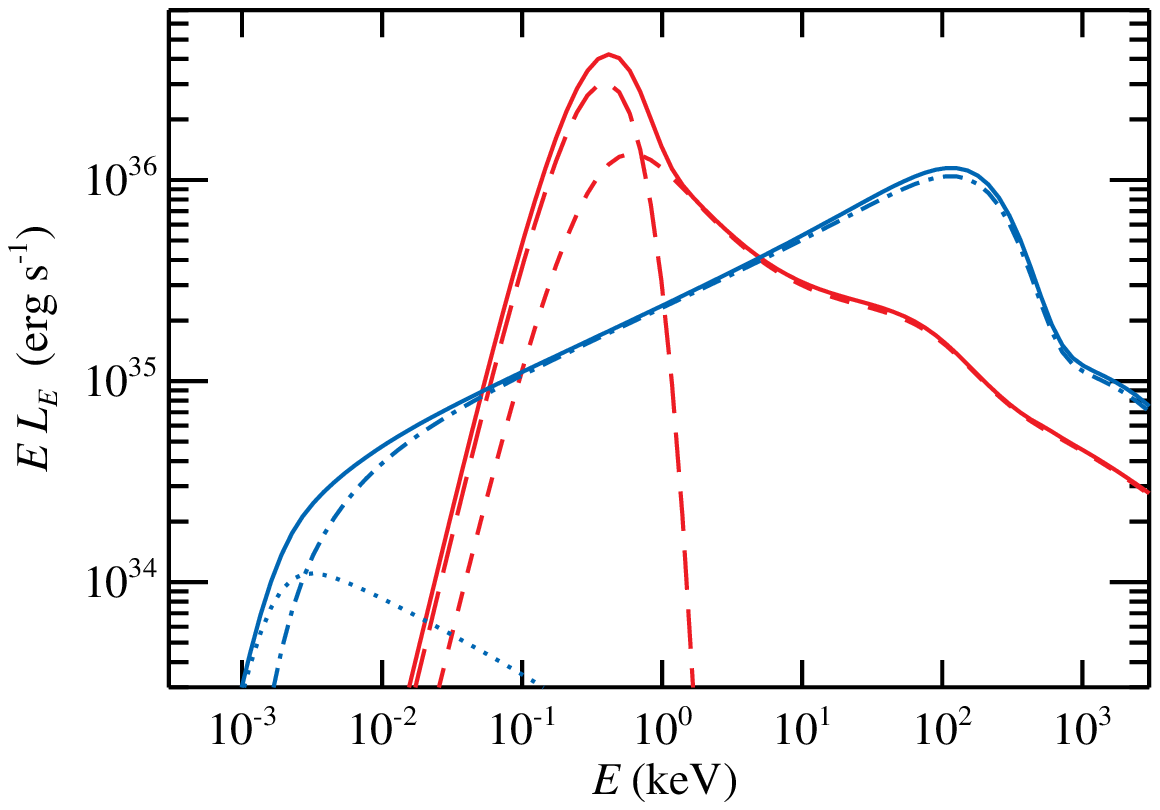}
\caption{
Left: schematic picture of the black hole vicinity: truncated cold accretion disc (red) and inner hot flow (blue).
$R_{\rm DC}$ and $R_{\rm SC}$ denote the characteristic radii of disc- and synchrotron Comptonization.
The fluctuations propagate from $R_{\rm DC}$ to $R_{\rm SC}$ on a timescale of $t_0$.
Right: average spectrum of the Comptonizing medium with two sourced of seed photons. 
Black line corresponds to the overall spectrum and different components are shown: synchrotron (green dotted line), synchrotron Comptonization (green dot-dashed line), disc blackbody (red long-dashed line) and disc Comptonization (red dashed line).
}\label{fig:geom} 
\end{figure*}

% (Fig. 2)
\begin{figure}
\centering 
\includegraphics[width=7cm]{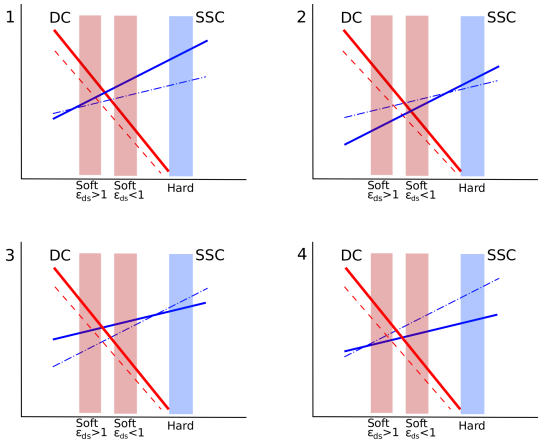}
\caption{
Schematic representation of the different behaviour of disc- and synchrotron Comptonization continua.
Thick solid lines correspond to the continua for the increased mass accretion rate. 
In all four panels we assume the disc Comptonization luminosity increases as the mass transfer rate increases. 
Cases 1 and 2 consider enhanced emission in hard X-rays in response to the increased mass accretion rate, while cases 3 and 4 consider the hard X-ray emission is decreased.
Cases 1 and 4 consider the regime when the pivoting point does not belong to the X-ray range, and cases 2 and 3 consider the pivoting point between the soft and hard energy bands.
}\label{fig:scheme} 
\end{figure}

Contradictions in understanding of spectral formation in X-ray binaries complicate the understanding of the nature of variability in these objects.
The short-term timing properties of accreting X-ray binaries have been investigated from the early 1970s \citep{Terrell72} and revealed the power spectra in the hard state can be described by a double broken power-law with $f^{0}$ ($f$ is the Fourier frequency) below $\sim10^{-1}$~Hz, $f^{-1}$ distribution in the range $\sim10^{-1}-10$~Hz and a steeper slope, roughly $f^{-2}$, at higher frequencies \citep{BH90}.
The wide range of frequencies demonstrated by these sources was not in line with the understanding that the X-rays are emitted predominantly from the vicinity of compact objects.
It was realised that the fluctuations are not excited in the same place where they are transformed to the X-ray photons, but rather they travel from the far regions inwards.
The propagating fluctuations model was proposed, whose essence is that the mass accretion rate fluctuations are excited in the disc at certain characteristic frequencies, and then they propagate towards the compact object, where the fluctuations at higher frequencies are excited \citep{Lyub97}. 

A major breakthrough in the field was made with the launch of the {\it Rossi X-ray Timing Explorer} ({\it RXTE}), when the systematic study of the evolution of timing properties throughout the outburst became possible.
Discovery of time lags between different X-ray energy bands \citep{MiKi89,NWD99,KCG01} and the linear relation between the root mean square (rms) variability amplitude and the X-ray flux \citep{UM01} allowed further establishment of the propagating fluctuations model.
In addition to the aperiodic variability, the {\it RXTE} capabilities allowed ubiquitous identification of the narrow features, the low-frequency quasi-periodic oscillations, in the hard and intermediate state power spectra.
Their frequency was found to evolve, in a correlated way, with the low-frequency break of the power-law component \citep{WvdK99,PBvdK99}.
The correlated changes were attributed to the change of the geometry during the state transition, when the standard accretion disc inner radius decreases as the source moves towards the soft state.

The exciting capabilities of the X-ray space observatory {\it XMM}$-$Newton allowed to explore spectral and timing properties in the energy range 0.1$-$2~keV.
Interconnection of soft (below $\sim$1~keV) and hard bands posed number of questions to the proposed picture of the ``stable disc and unstable corona'' \citep[proposed in][]{CGR01}, suggesting substantial short-term variability in the soft X-ray band, where the cold disc is thought to be dominant \citep[e.g.,][]{WU09,UWC11,CUM12}.
Lately, the model of propagating fluctuations was challenged by the analysis of the joint timing characteristics of soft and hard X-ray bands.
The long-term variability in the soft band was often seen to dominate over the variability in the hard band \citep{WU09,UWC11,RIvdKcygx1}.
Moreover, the soft band was sometimes found to lack power at low frequencies, as compared to the hard band \citep{RIvdK17j1550}.
This is counter-intuitive to the predictions of the propagation model, according to which the power spectra should be similar at low frequencies, and may differ at high frequencies, which in turn originate from the innermost region and are expected to be present only in the hard component.

The problems seem to occur during the intermediate states, when the spectra may be composed of two Comptonization continua.
It was found that the linear rms--flux relation no longer holds \citep{MDMB11}.
The shapes of the X-ray power spectra during the intermediate states are also peculiar.
A number of works quote presence of multiple peaks, as opposed to the simple $f^{-1}$ power-law \citep{NWD99,RTB00,Homan01,BPvdK02,Belloni06}.
These peaks can be produced by the interference of the two X-ray continua \citep{V16}, or by the propagating outwards fluctuations \citep{MIvdK17}, or be the result of an enhanced variability at certain radii in the flow, for instance, in the transitional region between the cold disc and hot flow or at hot flow inner boundary \citep{VCG94b,F09,ID12}.

A promising prospect for understanding of processes in X-ray binaries is the joint analysis of the spectral and timing characteristics. 
Several techniques have been applied, including reverberation modelling \citep{Pou02,DeMarco15reverber,DeMarco16reverber}, frequency resolved spectroscopy \citep{RGC99,GCR00,ADH14}, phase spectroscopy of the quasi-periodic oscillations \citep{AU16}, principal component analysis \citep{dSMJ08} and the joint fitting of time-averaged and Fourier-resolved spectra \citep{MD18a,MD18b}.
Most of the studies rely on the assumption of the coherent fluctuations in different bands, which can essentially be reduced to the variability of a normalization of the pre-assumed or fitted components.
Analysis of variability involving changes of spectral shape is currently the most challenging.
Lately, non-linear effects have been incorporated into the reverberation modelling \citep{MIvdK18}, studies of the long-term spectral evolution \citep{K15} and in multiwavelength modelling \citep{M14}.

In this work, we investigate timing properties of the light curves containing two spectral components. 
We incorporate non-linear effects into the short-term timing modelling in the context of propagating fluctuations.
In particular, the analysis includes the cases of pivoting power-law.
We study joint characteristics of the hard and soft X-ray band, such as cross-spectra, cross-correlation functions and phase lags.
We make predictions for the evolution of these characteristics in the course of the outburst and show that the presence of two components can resolve the challenges to the model of propagating fluctuations.
The results reiterate the importance of understanding the conditions of spectral formation and their evolution in explaining timing properties.

% (Fig. 3)
\begin{figure*}
\centering 
\includegraphics[width=14cm]{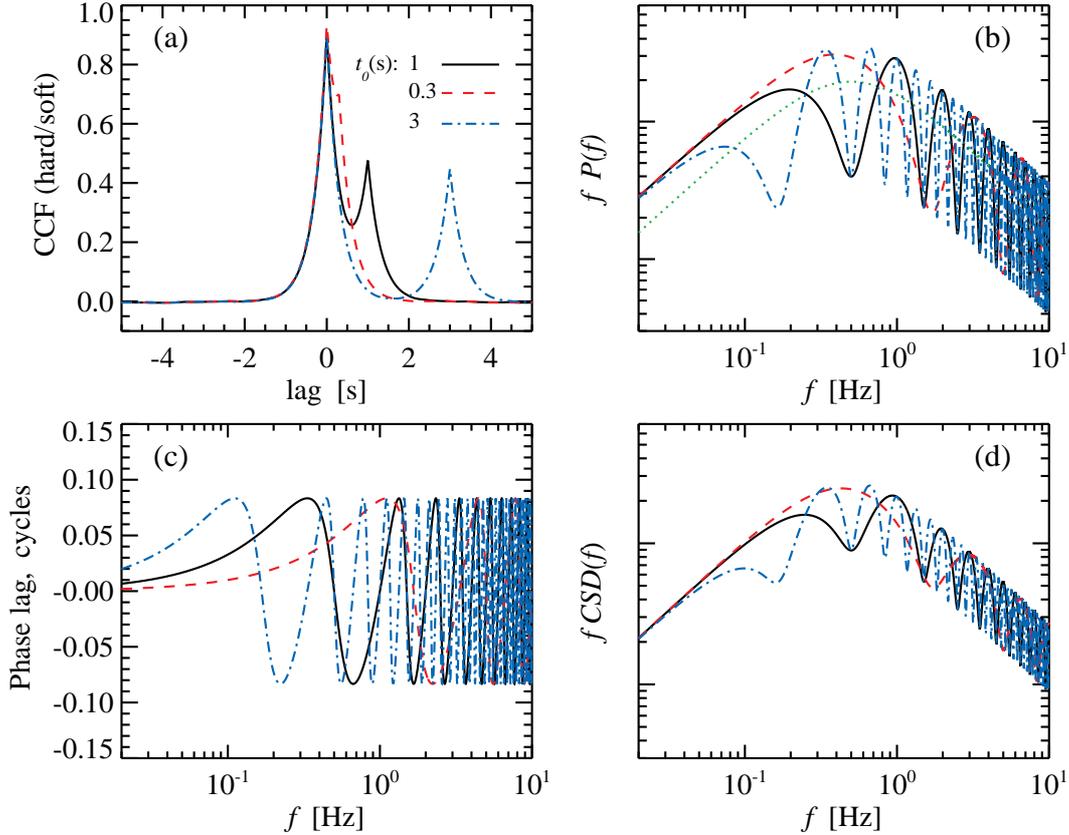}
\caption{
The cross-correlation functions (a), power spectra (b), phase lags (c) and cross-spectra (d) obtained for different delays $t_0$. 
Black solid lines correspond to $t_0=1$~s, red dashed $t_0=0.3$~s, blue dot-dashed $t_0=3$~s.
Green dotted line corresponds to the hard band power spectrum.
Other parameters are listed in Table~\ref{tab:par}.
}\label{fig:t0} 
\end{figure*}

% (Fig. 4)
\begin{figure*}
\centering 
\includegraphics[width=14cm]{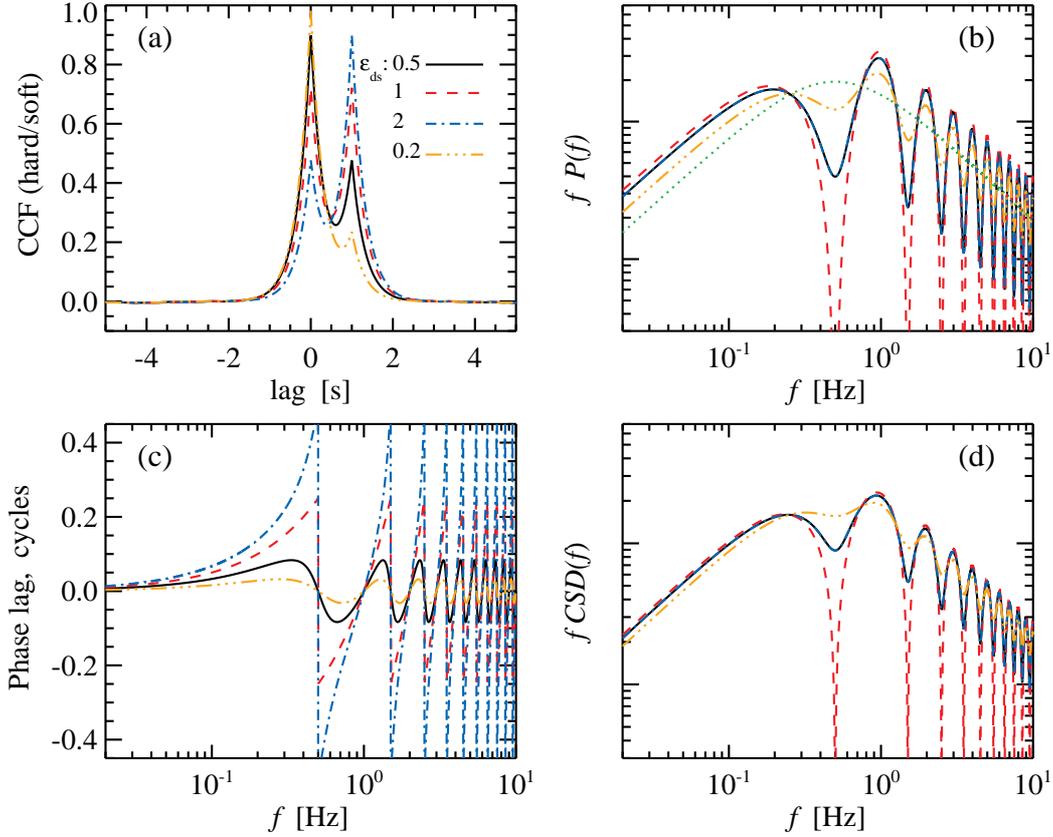}
\caption{
Same as in Fig.~\ref{fig:t0}, but for varying ratio of X-ray components $\varepsilon$.
The vertical lines in panel (c) mark frequencies where discontinuities occur.
All parameters are listed in Table~\ref{tab:par}.
}\label{fig:eps} 
\end{figure*}

% (Fig. 5)
\begin{figure*}
\centering 
\includegraphics[width=14cm]{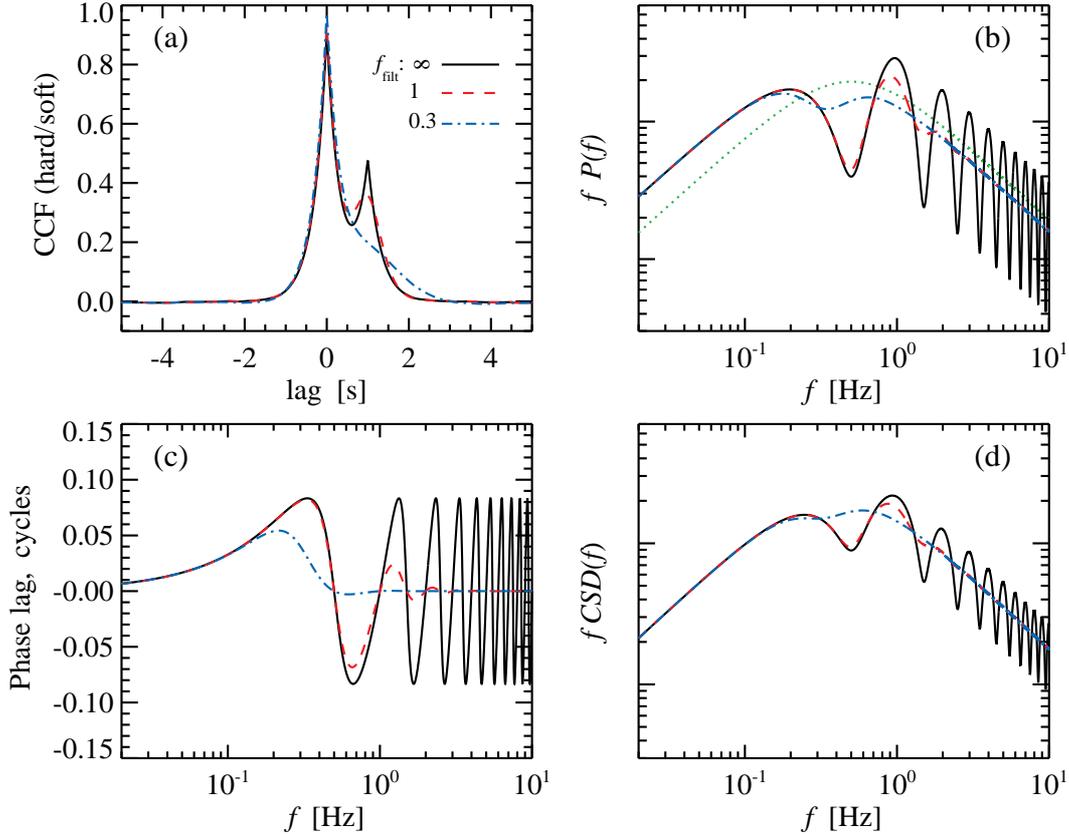}
\caption{
Same as in Fig.~\ref{fig:t0}, but for varying filtering frequency $f_{\rm filt}$.
All parameters are listed in Table~\ref{tab:par}.
}\label{fig:filt} 
\end{figure*}

% (Fig. 6)
\begin{figure*}
\centering 
\includegraphics[width=14cm]{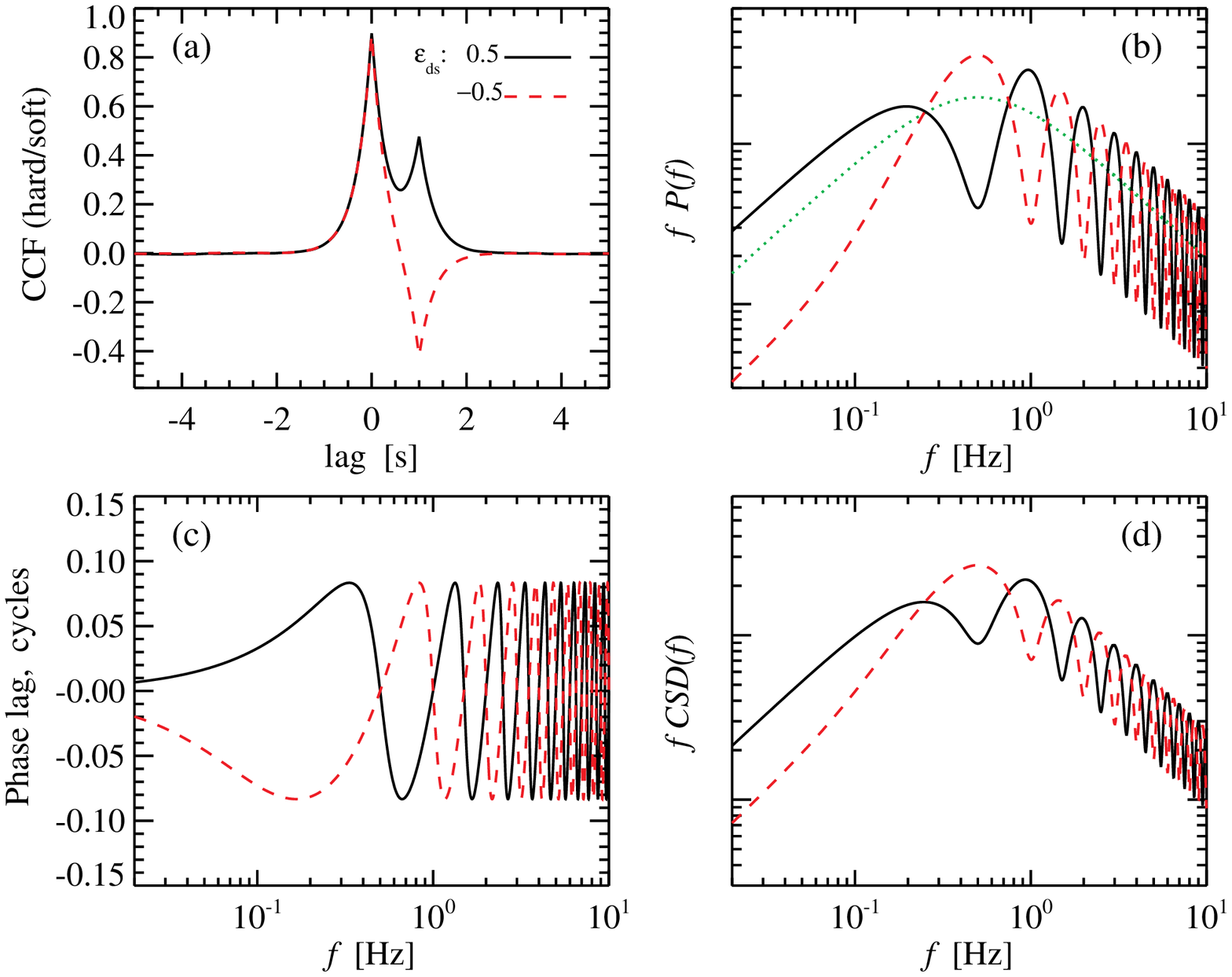}
\caption{
Same as in Fig.~\ref{fig:t0}, but for different signs of $\varepsilon_{\rm ds}$.
}\label{fig:anti} 
\end{figure*}

\section{Physical picture}

We develop a mathematical model for joint timing characteristics of two energy bands.
The model can be applied to any data with two spectral components in the light curves.
Though no crucial mathematical assumptions on the spectral components are made, the parameters acquire meaning only in the context of the physical scenario.
We consider the scenario of the truncated cold disc and the hot accretion flow within its inner boundaries \citep{PKR97,Esin97}.
The hot accretion flow corresponds to some type of radiatively inefficient solution \citep{KFM08,YN14}, which radiates in the X-rays via proceeding Comptonization processes.
The seed photons are coming both from the cold disc and the self-generated synchrotron emission \citep{PV14}.

We assume that the mass accretion rate fluctuations can be excited in the cold disc and/or in the hot flow and then propagate towards the compact object \citep{Lyub97,CGR01,AU06,IvdK13}.
The accretion rate fluctuations lead to variations of the seed photon flux in the cold disc, of the liberated energy and of the particle number density in the hot accretion flow. 
These variations lead to changes of the X-ray flux: first time, through the disc Comptonization in the region close to the disc truncation radius, and second time, through the 
synchrotron Comptonization in the region close to the black hole (within about $30-50$ Schwarzschild radii).
Schematic geometry and an example of an average spectrum\footnote{The spectrum was calculated in a spherical geometry, homogeneously filled with electron gas, which is accelerated through particle injection and cooled/thermalised by radiation. Two separate zones are considered: with the external disc photons and without them. The parameters of the medium are: radius of the sphere $3\times10^7$~cm, its total luminosity $5\times10^{36}$~erg~s$^{-1}$, disc blackbody temperature $0.1$~keV and luminosity of the injected photons $5\times10^{36}$~erg~s$^{-1}$, Thomson optical depth of the medium is 2 and the magnetic field is $3\times 10^5$G. The code was described in \citet{VP09,VVP11}.}
are shown in Fig.~\ref{fig:geom}.

The two X-ray continua may possess different hardness.
The X-ray variations arising from synchrotron Comptonization are delayed, with respect to the variations mediated through disc Comptonization, by the time it takes fluctuations 
to propagate from the truncation radius to the zone where synchrotron is emitted, which is of the order of viscous timescale in the hot accretion flow.
The detected X-ray light curve contains variations of these two sources, one is delayed with respect to another.
This leads to the interference picture in the X-ray power spectra \citep{V16}.

Let us consider the region where the disc photons are effectively Compton up-scattered.
The geometry in this case resembles the well-known slab corona model \citep{HM93}, which is known to give soft spectra \citep{SPS95,PVZ18}.
For high seed photon flux, the electrons are rapidly cooled, and for the coronal Thomson optical depth $\sim$1 attain temperatures below $50$~keV.
The spectrum achieves photon index $\Gamma\sim2.4$, hence the contribution of this component is large in the soft X-rays and is negligible in the hard X-rays.
On the contrary, the synchrotron Comptonization spectrum is typically hard \citep{PV09,VPV13}, with photon index $\Gamma\sim1.7$.
We then expect negligible to no contribution of the disc Comptonization component in the hard X-rays, and simultaneous presence of two components in the soft X-rays.
The light curves properties then depend on the considered spectral energy range.

On the other hand, the hot plasma atop of the cold accretion disc may contain substantial fraction of non-thermal particles.
The non-thermal population may arise from the proceeding particle acceleration, e.g., in the reconnection events of the buoyant magnetic loops \citep{SS14,B17}.
In this case, in addition to the soft thermal continuum, there is a non-thermal Comptonization continuum, which can be flat and extend up to MeV energies (similar to the soft-state spectrum of Cyg~X-1, \citealt{PC98,G99,PV09,MB09}).
In this case, there are two components in both hard and soft bands.

\section{Two components only in the soft band}

We start with the simplest case assuming that the disc Comptonization is mostly thermal and soft, and gives significant contribution only to the soft band.
The hard band is then produced solely by synchrotron Comptonization, while soft band consists of two components.
Though this is a special case of the more general scenario with two components in both soft and hard band, it is more intuitive and thus deserves a separate consideration.

\subsection{Mathematical formulation}

We assume that the light curves produced by both synchrotron and disc Comptonization are directly proportional to the mass accretion rate $\dot{m}(t)$ (see case 1 in Fig.~\ref{fig:scheme}).
The hard-band light curve is described as
\begin{equation}
 h(t) = \dot{m}(t).
\end{equation}
Its power spectrum is 
\begin{equation}\label{eq:psd_hard}
 P_{\rm h}(f) = \dot{M}^*(f) \dot{M}(f).
\end{equation}
Hereafter, we use capital letters to denote the Fourier transforms of the corresponding quantities in the time domain.
The soft component consists of two continua: one coming from the disc Comptonization and the other coming from the synchrotron Comptonization, both are proportional to the mass accretion rate, but the synchrotron Comptonization component is delayed with respect to the disc Comptonization component
\begin{equation}
 s(t) = \varepsilon_{\rm ds} \dot{m}(t+t_0) \ast g(t) + \dot{m}(t).
\end{equation}
Here, the first term corresponds to disc Comptonization and the second term corresponds to synchrotron Comptonization.
The ratio of two components in the soft band is regulated by the parameter $\varepsilon_{\rm ds}$ and $t_0$ is the relative delay. 
The ratio $\varepsilon_{\rm ds}$ has the meaning of the disc Comptonization to synchrotron Comptonization ratio of the absolute root mean squire variability amplitudes. 
It can be thought of as the ratio of areas between the upper and lower lines (at corresponding band) in Fig.~\ref{fig:scheme}.
Function $g(t)$ is the low-pass filter, it is responsible for damping of the high-frequency fluctuations in the disc Comptonization.
Its Fourier transform is considered in the form
\begin{equation}
 G(f) = \frac{1}{(f/f_{\rm filt})^4+1},
\end{equation}
where $f_{\rm filt}$ gives the characteristic damping frequency.

We consider joint characteristics of the soft and hard components: power spectra, cross-spectral amplitudes and phase lags and cross-correlation function.
The power spectra of soft $P_{\rm s}(f)$ and hard $P_{\rm h}(f)$ components are related as (see also \citealt{V16})
\begin{equation}\label{eq:psd_soft}
 P_{\rm s}(f) = P_{\rm h}(f) \left[ 1 + \varepsilon_{\rm ds}^2G^2(f) + 2\varepsilon_{\rm ds} G(f)\cos{(2\pi f t_0)} \right].
\end{equation}
For $\varepsilon_{\rm ds}>0$, the power spectrum of the soft component demonstrates oscillations with peaks appearing at frequencies $f=k /t_0$ (where $k$ is an integer number), and the power is significantly reduced around frequencies $f=(k-1/2)/t_0$.

The cross-spectral density can be calculated as
\begin{equation}
 CS(f) = S^*(f) H(f) = P_{\rm h}(f) \left[ \varepsilon_{\rm ds} e^{2 \pi i f t_0} G(f) + 1  \right].
\end{equation}
The cross-amplitudes are related to power spectra as
\begin{equation}\label{eq:cs_ampl}
 |CS(f)| = P_{\rm h}(f)  \biggl\{1 + \varepsilon_{\rm ds}^2G^2(f) + 2\varepsilon_{\rm ds} G(f)\cos{(2\pi f t_0)}  \biggl\}^{\!1/2}.
\end{equation}
Hence, the cross-spectral density is expected to show peaks similar to the power spectrum of the soft component.

The phase lag spectrum $\Delta \varphi (f)$ can be found from relation
\begin{equation}\label{eq:phase}
 \tan \Delta \varphi (f) = \frac{\varepsilon_{\rm ds} G(f) \sin (2 \pi f t_0) }{1 +  \varepsilon_{\rm ds} G(f) \cos (2 \pi f t_0)}.
\end{equation}
The phase lags are defined so that positive values correspond to lags of the hard component.
Similar to the power spectrum and cross-spectrum, the phase lags demonstrate oscillatory behaviour, with the peaks appearing at frequencies
\begin{equation} 
  f=\frac{k}{t_0} \pm \frac{\pi - \arccos\varepsilon_{\rm ds}}{2\pi t_0}
\end{equation}
for $\varepsilon_{\rm ds}<1$ and $G(f)=1$.
In contrast to the power spectra, the position of the phase lags peaks depends on the ratio $\varepsilon_{\rm ds}$.
The maximum/minimum value of the phase lag are defined as
\begin{equation}
\Delta \varphi_{\rm max,min} = \pm \arctan \frac{\varepsilon_{\rm ds}}{\sqrt{1-\varepsilon_{\rm ds}^2}}.
\end{equation}

For values $\varepsilon_{\rm ds}\geq1$, the phase lags are monotonic functions of frequency.
We determine the phase lags in the range $[-\pi; \pi)$, and frequencies at which $\Delta \varphi = \pi$ is achieved depend on the delay as $\displaystyle f=(k-1/2)/t_0$.

No simple analytical expression describing the shape of the cross-correlation function can be found in a general case.
For the case of no filtering, the resulting CCF consists of a sum of two auto-correlation functions of mass accretion rate, one of them shifted and multiplied by $\varepsilon_{\rm ds}$:
\begin{equation}
{\rm  CCF} (\Delta t) \propto {\rm ACF}_{\dot m} (\Delta t) + \varepsilon_{\rm ds} {\rm ACF}_{\dot m} (\Delta t - t_0).
\end{equation}
The CCFs are expected to demonstrate a double-peak structure due to presence of two terms in the soft component.
Two peaks, however, can be wide enough to overlap, and a skewed single-peak CCFs in expected in this case.

\begin{table}
\caption{Parameters of numerical modelling for Figs~\ref{fig:t0}--\ref{fig:sign}. The sign in the last column corresponds to the correlation/anti-correlation of hard and soft synchrotron Comptonization light curves with $\dot{m}(t)$, respectively.
}\label{tab:par}
  \begin{center}
\begin{tabular}{ccccccc}
\hline
\hline
Figure 	& Line			&	$t_0$~(s)	& $\varepsilon_{\rm ds}$	&$f_{\rm filt}$~(Hz) & {\rm sign}\\
\hline 
\ref{fig:t0},\ref{fig:eps},\ref{fig:filt},\ref{fig:anti},\ref{fig:sign}	& black solid		&	1		&	0.5		&	$\infty$	& $+$ $+$\\
\ref{fig:t0}										& red dashed		&	0.3	 	&      0.5   		&	$\infty$	 & $+$ $+$ \\
\ref{fig:t0}										& blue dot-dashed	&	3.0	 	&      0.5   		&	$\infty$	  &$+$ $+$\\
\ref{fig:eps}									& red dashed		&	1	 	&      1   		&	$\infty$	  &$+$ $+$\\
\ref{fig:eps}									& blue dot-dashed	&	1	 	&      2    		&	$\infty$	  &$+$ $+$\\
\ref{fig:eps}									& yellow triple dot-dashed	&	1	&      0.2   		&	$\infty$	  &$+$ $+$\\
\ref{fig:filt}										& red dashed		&	1	 	&      0.5   		&	1.0	          &$+$ $+$\\
\ref{fig:filt}										& blue dot-dashed	&	1	 	&      0.5   		&	0.3	          &$+$ $+$\\
\ref{fig:anti}  									& red dashed		&	1		&      $-$0.5	&	$\infty$	 & $+$ $+$\\
\ref{fig:sign}  									& green dotted		&	1		&      0.5	&	$\infty$	& $+$ $-$  \\
\ref{fig:sign}  									& red dashed		&	1		&      0.5	&	$\infty$	& $-$ $+$   \\
\ref{fig:sign}  									& blue dot-dashed		&	1		&      0.5	&	$\infty$	& $-$ $-$   \\
\hline
     \end{tabular}
  \end{center}
\end{table}

\subsection{Examples}

The dependence of the power and cross-spectra and phase lags on model parameters can be traced using the analytical expressions in Eqs.~(\ref{eq:psd_soft}), (\ref{eq:cs_ampl}) and (\ref{eq:phase}).
We use results of numerical simulations to demonstrate the CCFs for different parameters.

We compute the light curves in different bands from their power spectral densities (PSDs).
We start with the determination of the mass accretion rate power spectrum.
For an illustration, we assume it to be in the form of a zero-centred Lorentzian function
\begin{equation}
 |\dot{M}|^2(f)={\cal L}(f)\equiv\frac{r^2}{\pi} \frac{\Delta f }{(\Delta f)^2 + f^2}, 
\end{equation}
where $r$ gives the total rms and $\Delta f$ is the width. 
We consider $\Delta f = 0.5$~Hz (this peak frequency is within the range of frequencies the observed variability peaks at) and leave $r$ as free parameter.
Once the shape of the hard X-ray power spectrum is fixed, three parameters are left: $t_0$, $\varepsilon_{\rm ds}$ and $f_{\rm filt}$.
The fiducial values are chosen as $t_0=1$~s, $\varepsilon_{\rm ds}=0.5$, $f_{\rm filt}\to \infty$.
We study how the timing characteristics change under the variations of these three parameters.
The resulting CCFs (panels a), PSDs (panels b), phase lags (panels c) and cross-spectral amplitudes (panels d) are shown in Figs~\ref{fig:t0}--\ref{fig:filt} and the parameters are listed in Table~\ref{tab:par}.

% (Fig. 7)
\begin{figure*}
\centering 
\includegraphics[width=14cm]{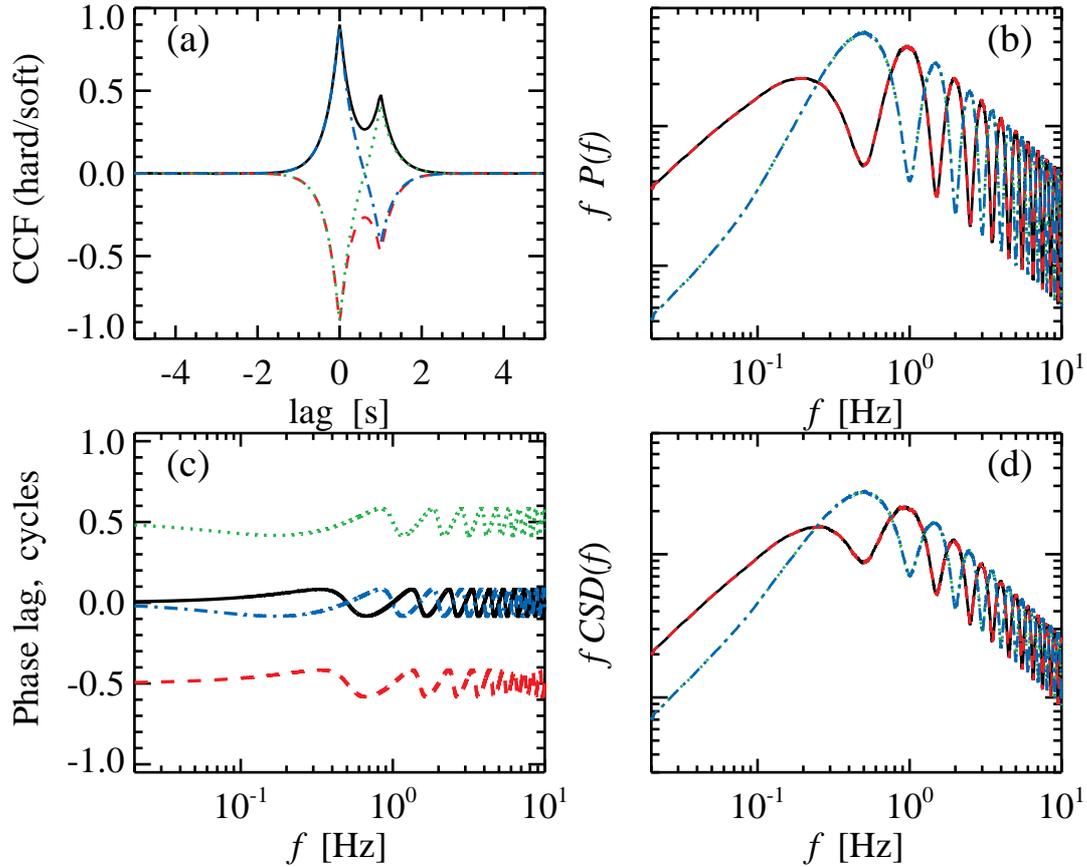}
\caption{
Comparison of the timing characteristics for the cases illustrated in Fig.~\ref{fig:scheme}: correlation of both synchrotron and disc Comptonization with $\dot{m}(t)$ with fiducial parameters (case 1, black solid line), correlation of the hard-band synchrotron Comptonization component and anti-correlation of the soft band with $\dot{m}(t)$ (case 2, green dotted line), correlation of synchrotron Comptonization soft X-rays and anti-correlation of hard X-rays with $\dot{m}(t)$ (case 3, red dashed line) and anti-correlation of both hard and soft X-rays with $\dot{m}(t)$ (case 4, blue dot-dashed line).
}\label{fig:sign} 
\end{figure*}

% (Fig. 8)
\begin{figure}
\centering 
\includegraphics[width=7cm]{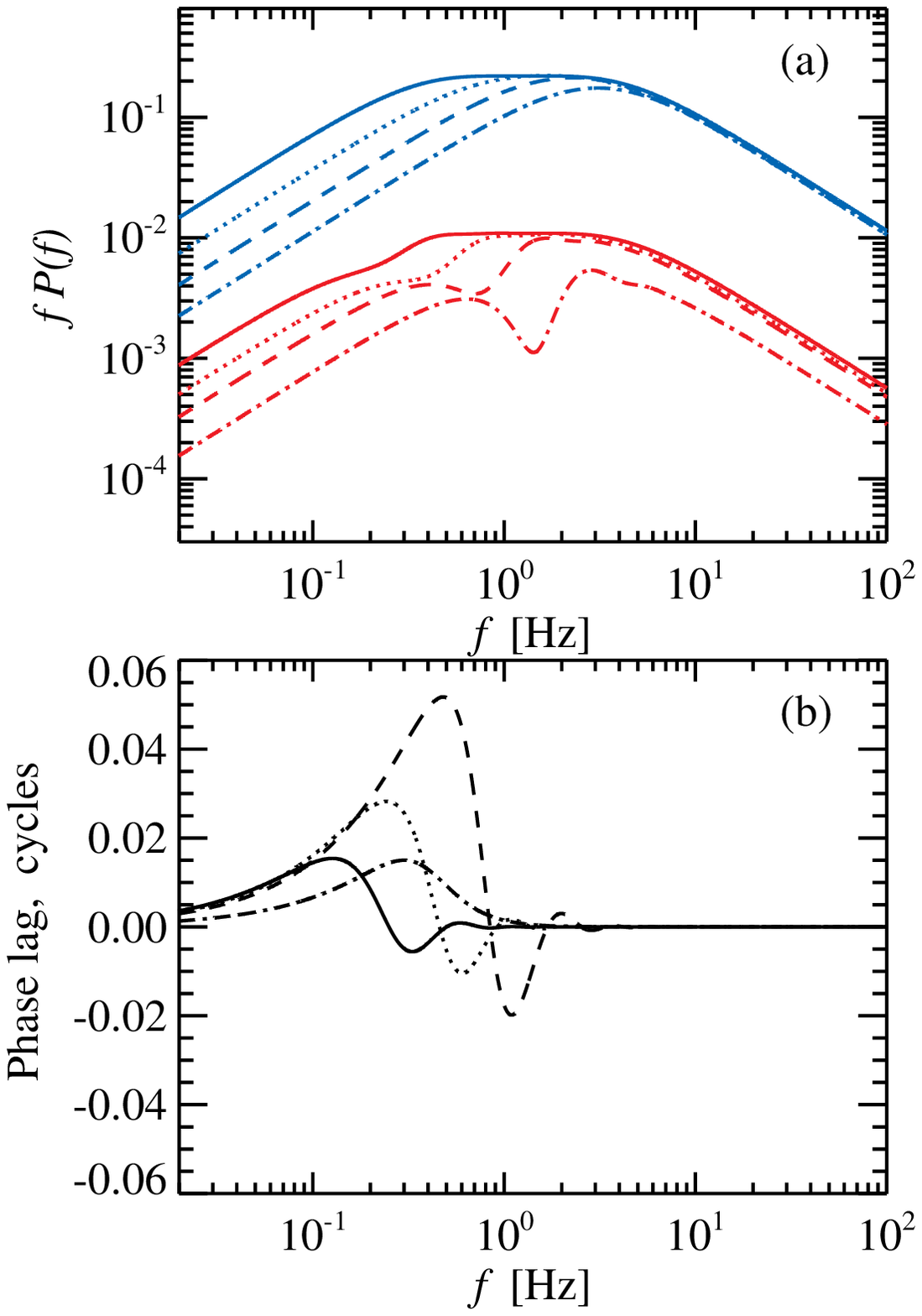}
\caption{
Evolution of power spectra (a) and phase lags (b) during the state transition.
The upper curves in panel (a) correspond to the hard band and the lower curves correspond to the soft band.
The solid line corresponds to the hardest spectrum and largest hot accretion flow, and the dot-dashed line corresponds to the smallest truncation radius.
Parameters are listed in Table~\ref{tab:trans} and the normalisations of PSDs are arbitrary.
}\label{fig:trans} 
\end{figure}

The increase of the time delay $t_0$ (Fig.~\ref{fig:t0}) is reflected in a more pronounced two-peak structure of the CCF.
For the delays comparable to the width of the accretion rate ACF, the CCF is one-peak and asymmetric. 
The power spectrum of the soft component, cross-spectrum and phase lag spectrum demonstrate more peaks as $t_0$ increases.
The position of the first peak increases with increasing delay in the phase lags, but at the same time it decreases in the power- and cross-spectra.
We note an artificial broadening of the overall soft-band PSD (with two spectral components), as compared to the hard band (with one spectral component).
The highest power, in general, does not correspond to the peak of the original Lorentzian.

The change of the disc to synchrotron Comptonization ratio $\varepsilon_{\rm ds}$ leads to redistribution of the peak amplitudes in the CCF (see Fig.~\ref{fig:eps}).
The oscillations are most pronounced in the power spectra for $\varepsilon_{\rm ds}=1$.
The power and cross-spectra are identical for $\varepsilon_{\rm ds}$ and $1/\varepsilon_{\rm ds}$, however, the phase lags are not.
The peaks in phase lags become narrower as $\varepsilon_{\rm ds}$ approaches unity and in the data will likely be smeared out by the averaging with neighbouring frequencies and by variations of parameters.
For values  $\varepsilon_{\rm ds}\geq1$, the phase lag spectra are increasing function of frequency (see vertical lines in Fig.~\ref{fig:eps}c marking the frequencies where $\Delta \varphi = \pi$).

In Fig.~\ref{fig:filt} we demonstrate the dependence on damping frequency, which is rather straightforward.
The smaller is the frequency $f_{\rm filt}$, the more smeared the CCF is, and for sufficiently low frequency it loses the secondary peak (around time lag 1~s in Fig.~\ref{fig:filt}a).
The power and cross-spectra and the phase lags demonstrate less peaks, as at frequencies $f>f_{\rm filt}$ only one component dominates, and hence no interference appears.

\subsection{Pivoting points and alteration of timing characteristics}

The consideration above concerns the case when variations in mass accretion rate cause variations in Comptonization spectrum of disc and synchrotron photons of the same sign.
The situation can be realised if, e.g. the increase of accretion rate leads to higher energy dissipation per particle, leading to harder spectra and higher luminosity.
However, in the case when the amount of seed photons in turn depends on the accretion rate, the increase of the latter might result in spectral softening at higher X-ray luminosity.
The ``harder when brighter'' is naturally realised in the case of synchrotron Comptonization spectra \citep[e.g., fig.~1 of][]{VPV11}, while the ``softer when brighter'' behaviour is  expected in the case of disc Comptonization.
Both of these patterns were detected in the short-term X-ray variations \citep{MBSK03,GMD08}, as well as on the long timescales \citep{ZPP02}.
The location of the pivoting point in synchrotron Comptonization spectrum depends on the parameter which causes major changes of spectral shape (optical depth, magnetic field etc., see fig.~1 of \citet{PV09}).
However, because of the absence of understanding of how the changes in mass transfer rate are reflected in the changes of system parameters, there are numerous options of how the spectra react to the increased $\dot{m}(t)$.
Likewise, there is no unique solution for the disc Comptonization spectra and their pivoting points.
Below we consider alternative possibilities for the behaviour of hard and soft light curves in response to the fluctuating $\dot{m}(t)$.

We first investigate how the timing characteristics change if we change the sign of $\varepsilon_{\rm ds}$.
This describes the case when the increase of mass accretion rate leads to a decrease of the disc Comptonization component in the soft X-rays (e.g., due to an excess cooling of the Comptonizing medium), but at the same time causes an increase of the synchrotron Comptonization in both energy bands.
The resulting timing characteristics are plotted in Fig.~\ref{fig:anti}.
The CCF shows the zig zag shape, similar to those obtained for anti-correlated optical synchrotron and Compton up-scattered X-ray emission \citep{VPV11}.
The power and cross-spectra demonstrate the change of dips into peaks and peaks into dips.
We also note a decrease of power at lower frequencies, at which fluctuations from disc and synchrotron Comptonization come in anti-phase and partially cancel each other.
The phase lag spectrum also demonstrates transformation of dips into peaks, however, the peaks for the case of anti-correlation do not exactly match the dips of the correlation case.
The maxima and minima now appear at 
\begin{equation}
 f=\frac{k}{t_0} \pm \frac{\arccos  |\varepsilon_{\rm ds}|}{2\pi t_0}
\end{equation}
for $-1<\varepsilon_{\rm ds}<0$.

We proceed with the case when the mass accretion rate variations lead to a decrease of the synchrotron Comptonization in both hard and soft X-rays, while the disc Comptonization is increased (see case 4 in Fig.~\ref{fig:scheme}).
This may happen if the increase of accretion rate causes substantial increase of the synchrotron photons, which lead to higher cooling and softer spectra. 
The light curves are then described as
\begin{eqnarray}
  h(t) &=& - \dot{m}(t) \\
  s(t) &=& \varepsilon_{\rm s} \dot{m}(t+t_0) \ast g(t) - \dot{m}(t),
\end{eqnarray}
and the signs in phase lags change accordingly.
The resulting CCF, phase lags, PSDs and cross-spectra are shown in Fig.~\ref{fig:sign} with blue dot-dashed lines.
The parameters and corresponding signs for hard and soft bands of the synchrotron Comptonization component are listed in Table~\ref{tab:par}.
The phase lags, power- and cross spectra are the same as in case of positive correlation of synchrotron Comptonization with accretion rate in both bands, but when using $\varepsilon_{\rm ds}<0$ (see Fig.~\ref{fig:anti}).

\begin{table}
\caption{Parameters of state transition modelling (Fig.~\ref{fig:trans}).
}\label{tab:trans}
  \begin{center}
\begin{tabular}{cccccc}
\hline
\hline
  Line 		&	$t_0$~(s)	& $\varepsilon_{\rm ds}$	&$f_{\rm filt}$~(Hz) & $\Delta f_1$~(Hz) & $\Delta f_2$~(Hz) 	\\
\hline 
 solid			&	2		&	0.1				&	0.3	 	        & 0.5				&  3.1	\\
 dotted		&	1.1	 	&      0.18 				&	0.55		        &  0.92 			& 3.1 \\
 dashed		&	0.6	 	&      0.34   			&	1	  		& 1.7				& 3.1 \\
 dot-dashed	&	0.32	 	&      0.62    			&	1.9	  		& 3.1				& 3.1 \\
\hline
     \end{tabular}
  \end{center}
\end{table}

We further consider the cases when the synchrotron Comptonization continuum pivoting point is located in the X-ray range, somewhere between the soft and hard X-rays (see cases 2 and 3 in Fig.~\ref{fig:scheme}).
The light curves can then be written as
\begin{eqnarray}
  h(t) &=& \pm \dot{m}(t) \\
  s(t) &=& \varepsilon_{\rm s} \dot{m}(t+t_0) \ast g(t) \mp \dot{m}(t).
\end{eqnarray}
The resulting timing characteristics are plotted in Fig.~\ref{fig:sign} with green dotted (case 2) and red dashed (case 3) lines.
The phase lags are located near the $\pm\pi$ points, but otherwise their dependence on frequency resembles that for the case 1 (correlation of both hard- and soft-band synchrotron Comptonization light curves with mass accretion rate), with $\varepsilon_{\rm ds}<0$ (case 2) and $\varepsilon_{\rm ds}>0$ (case 3).
The shapes  of PSDs and cross-spectra are the same as those in Fig.~\ref{fig:anti}: green line overlaps with the blue lines, and the red lines overlap with the black lines.

\subsection{Changes of timing properties during the transition}

During the state transition, all the parameters $\varepsilon_{\rm ds}$, $t_0$, $f_{\rm filt}$, as well as the shape of the mass accretion rate fluctuations power spectrum experience coordinated changes.
At the hard to soft transition, the low-frequency break of the mass accretion rate fluctuations power spectrum should increase if the fluctuations are produced in the shrinking hot accretion flow.
This causes the increase of the low-frequency break in the hard band power spectrum reported in a number of sources \citep[e.g., review][]{DGK07}.
In the soft band, we expect the role of disc Comptonization determined by $\varepsilon_{\rm ds}$ and the damping frequency $f_{\rm filt}$ to increase and the delay between the components $t_0$ to decrease. 
We consider the initial power spectrum in the hard band to be composed of two zero-centred Lorentzians, $\Delta f_1=0.5$~Hz and $\Delta f_2=3.1$~Hz, with equal power (see upper solid line in Fig.~\ref{fig:trans}a).
The characteristic widths determine the frequencies at which maximum variability power is concentrated, likely corresponding to characteristic viscous frequencies\footnote{The variability can be excited at frequencies between Keplerian and local viscous through the local dynamo processes, however, as the fluctuations propagate inwards in a diffusive way, frequencies higher than local viscous are suppressed \citep[see][]{HR15,MIvdK17}.} 
in the accretion flow (for simplicity, we consider only the fluctuations produced in the hot flow).
The power spectral shape resembles the double-broken power law.
We simulate the transition by simultaneously changing the parameters $t_0$, $\varepsilon_{\rm ds}$, $f_{\rm filt}$ and $\Delta f_1$ (see Table~\ref{tab:trans}), governed by the decrease of truncation radius.

The parameters depend on the truncation radius as follows. 
The propagation time is of the order of viscous timescale, hence $t_0 \propto R^{-3/2}$ (assuming constant aspect ratio and viscosity). 
The low-frequency Lorentzian width and the damping frequency are related to the timescales at the truncation radius, i.e. $f_{\rm filt} \propto R^{3/2}$ and  $\Delta f_1 \propto R^{3/2}$.
The dependence of the ratio of the synchrotron and disc Comptonization on truncation radius is not straight forward, however, it is clear that the ratio should increase with decreasing truncation radius.
We choose $\varepsilon_{\rm ds}\propto R^{-3/2}$ as an illustration and do not account for possible changes of sign of $\varepsilon_{\rm ds}$ at the transition.

We show one case of evolution of timing properties during the state transition as an illustration and leave the detailed fitting of the data to a future work.
We find that the interference pattern is not very prominent in the power spectra, as for high $t_0$, when the oscillations are apparent at low frequencies, we expect low damping frequency. 
As the damping frequency increases, the delay decreases, shifting the frequency of the first peak towards higher $f$.
The phase lags in the range 0.1$-$1~Hz first increase, and then decrease at the later stages of transition for the same reason (Fig.~\ref{fig:trans}b).
Thus, the average over the frequency domain phase lag does not necessarily have to be a monotonic function of the truncation radius.
We note that in contrast to the monotonic increase of the low-frequency break in the hard band, the position of the break in the soft band is not directly connected to the width of the low-frequency Lorentzian $\Delta f_1$.

\section{Two components in the hard and in the soft bands}

In this section, we develop the formalism for the more general case when both disc and synchrotron Comptonization components significantly contribute to the soft and hard bands.
In this case, we expect an interference picture in the power spectra of both energy ranges, but in addition, the cross spectra and the phase lags are more complex.

\subsection{Mathematical model}

We begin with the case when both hard and soft synchrotron Comptonization light curves are positively correlated with the mass accretion rate (case 1 in Fig.~\ref{fig:scheme})
The total hard and soft band light curves can be described as
\begin{eqnarray}
 h(t) &=& \varepsilon_{\rm h} \dot{m}(t+t_0) \ast g(t) + \dot{m}(t) \\
 s(t) &=& \varepsilon_{\rm s} \dot{m}(t+t_0) \ast g(t) + \dot{m}(t),
\end{eqnarray}
where the relative importance of the two continua are denoted as $\varepsilon_{\rm s}$ and $\varepsilon_{\rm h}$ in the soft and in the hard band, respectively.
The cross-spectral amplitudes are given by
\begin{equation}\label{eq:cs_ampl_two}
\begin{split}
 |CS(f)|&= P_{\rm  \dot{m}}(f)  \biggl\{
 \left[ 1 + \varepsilon_{\rm s}  \varepsilon_{\rm h} G^2(f) + (\varepsilon_{\rm s} + \varepsilon_{\rm h}) G(f) \cos{(2\pi f t_0)} \right]^2 \\
 &+  (\varepsilon_{\rm s} - \varepsilon_{\rm h})^2 G^2(f) \sin^2{(2\pi f t_0)} 
  \biggr\}^{\!1/2},
\end{split}
\end{equation}
where $P_{\rm  \dot{m}}(f)$ is the power spectrum of mass accretion rate fluctuations.
The phase lags can be calculated as
\begin{equation}\label{eq:phase_lags_two}
 \tan \Delta \varphi = \frac {(\varepsilon_{\rm s} - \varepsilon_{\rm h})G(f) \sin{(2\pi f t_0)}}
                                     {1 + \varepsilon_{\rm s}\varepsilon_{\rm h}G^2(f) + ( \varepsilon_{\rm s} + \varepsilon_{\rm h}) G(f) \cos{(2\pi f t_0)}},
\end{equation}
which reduces to Eq.~(\ref{eq:phase}) for $\varepsilon_{\rm h}=0$.
Interestingly, the sign of the phase lags at low frequencies depends on the sign of $(\varepsilon_{\rm s} - \varepsilon_{\rm h})$, and it can be positive even if both coefficients are negative.

There is a special case when $\varepsilon_{\rm s}=\varepsilon_{\rm h}$, which gives zero phase lags between two energy bands, even though one of the components is delayed with respect to another.
For the cases when the phase lags do not reach $\pm\pi$ (or, equivalently, when the derivative of the phase lag turns zero at some frequencies), their maxima/minima occur at frequencies (we put $G(f)=1$)
\begin{equation} \label{eq:freq_maxmin_two}
 f=\frac{k}{t_0} \pm \frac{\pi - \arccos\frac{\varepsilon_{\rm s}+\varepsilon_{\rm h}}{1+\varepsilon_{\rm s}\varepsilon_{\rm h}}}{2\pi t_0}.
 \end{equation}
When $|\varepsilon_{\rm s}|<1$ and $|\varepsilon_{\rm h}|<1$, the maximal/minimal value of the phase lag is
\begin{equation}\label{eq:deltaphimax_two}
 \Delta \varphi_{\rm max,min} = \pm \arctan \frac{\varepsilon_{\rm s}-\varepsilon_{\rm h}}{\sqrt{(1-\varepsilon_{\rm s}^2)(1-\varepsilon_{\rm h}^2)}}.
\end{equation}
When both $|\varepsilon_{\rm s}|>1$ and $|\varepsilon_{\rm h}|>1$, the minima and maxima are the same as for the case with $1/\varepsilon_{\rm s}$ and $1/\varepsilon_{\rm h}$, given that the denominator of Eq.~(\ref{eq:phase_lags_two}) is positive.
If the denominator is negative, the phase lags oscillate around $\pi$ with an amplitude $\Delta \varphi_{\rm max,min}$.

For the cases when the phase lags are increasing/decreasing function of frequency (the derivative does not turn zero), the frequencies at which $\Delta \varphi$ reaches $\pi$, are $\displaystyle f=(k-1/2)/t_0$ for 
$(\varepsilon_{\rm s}-\varepsilon_{\rm h})>0$, and $f=k/t_0$ for $(\varepsilon_{\rm s}-\varepsilon_{\rm h})<0$. 
The latter can be deduced by analysing the signs of the numerator and denominator in Eq.~(\ref{eq:phase_lags_two}).

% (Fig. 9)
\begin{figure}
\centering 
\includegraphics[width=7cm]{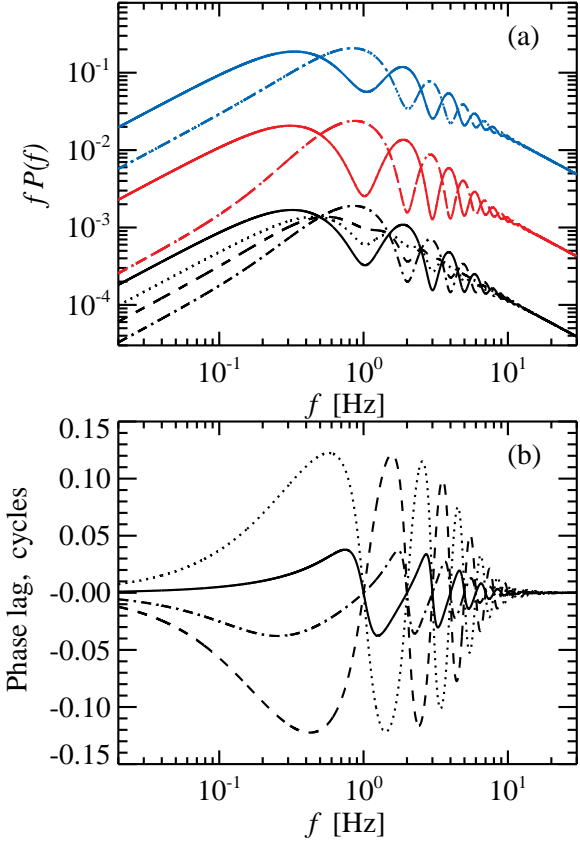}
\caption{
Power and cross-spectra (a) and phase lags (b) for the model with two components in both soft and hard X-rays demonstrating dependence on the signs of $ \varepsilon_{\rm s}$ and 
$ \varepsilon_{\rm h}$.
The upper (blue) curves in panel (a) correspond to the hard band, the (red) curves in the middle correspond to the soft band and the lower (black) curves correspond to the cross-spectra.
Parameters are listed in Table~\ref{tab:twocomp}.
}\label{fig:twocomp} 
\end{figure}

% (Fig. 10)
\begin{figure}
\centering 
\includegraphics[width=7cm]{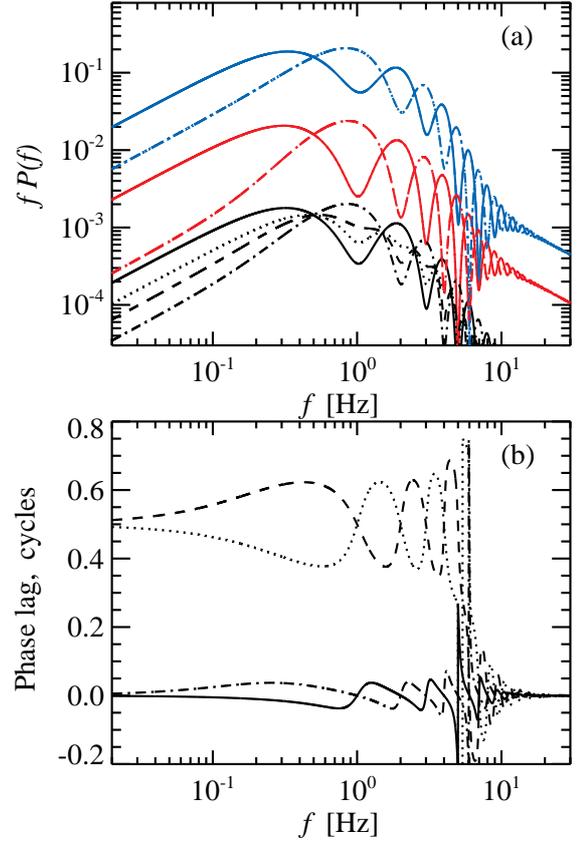}
\caption{
Same as in Fig.~\ref{fig:twocomp}, but for the inverse values of $ \varepsilon_{\rm s}$ and $ \varepsilon_{\rm h}$.
Parameters are listed in Table~\ref{tab:twocomp}.
}\label{fig:twocomp2} 
\end{figure}

% (Fig. 11)
\begin{figure}
\centering 
\includegraphics[width=7cm]{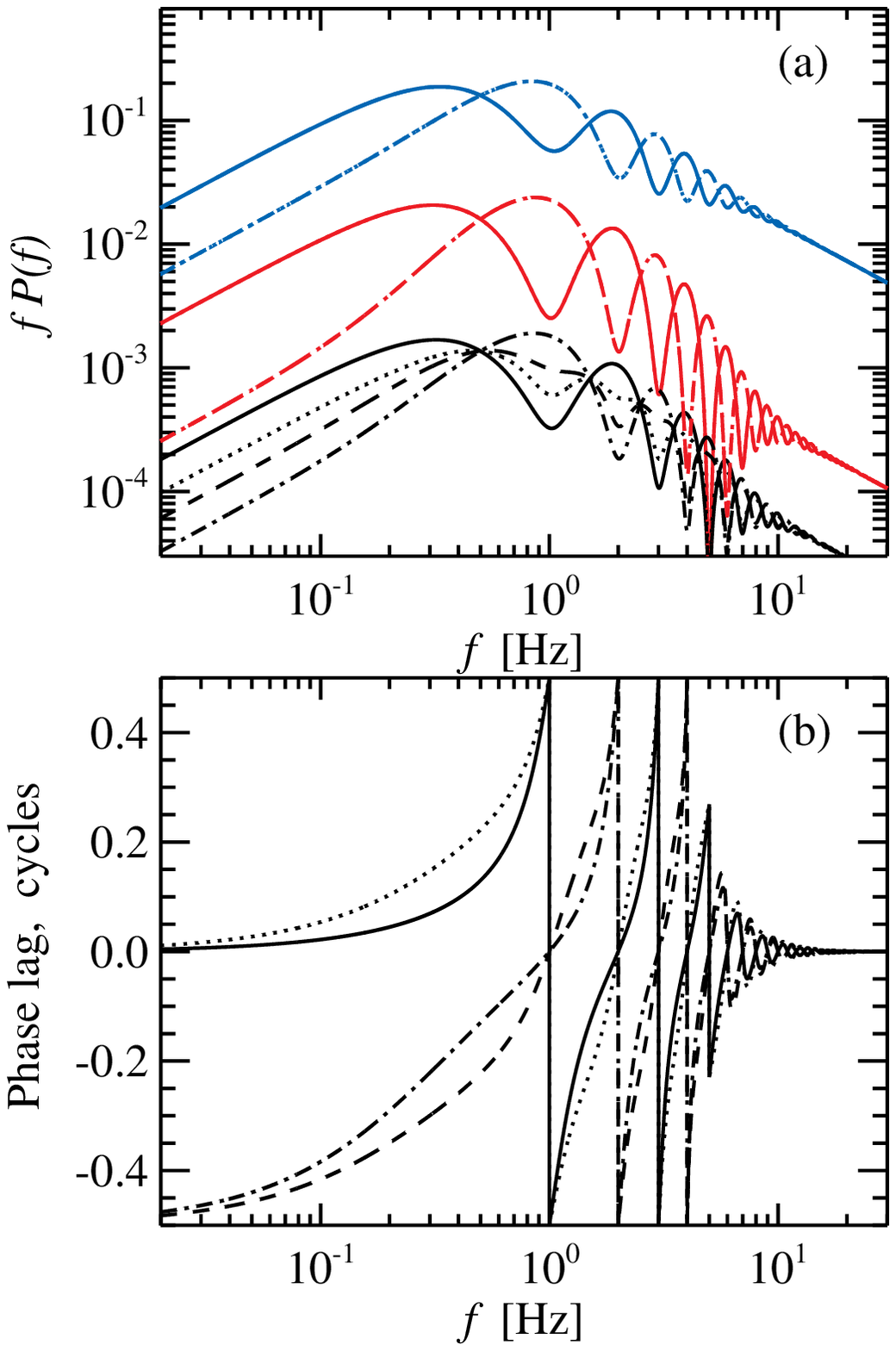}
\caption{
Same as in Fig.~\ref{fig:twocomp}, but for the inverse values of $ \varepsilon_{\rm h}$.
Parameters are listed in Table~\ref{tab:twocomp}.
}\label{fig:twocomp3} 
\end{figure}

% (Fig. 12)
\begin{figure}
\centering 
\includegraphics[width=7cm]{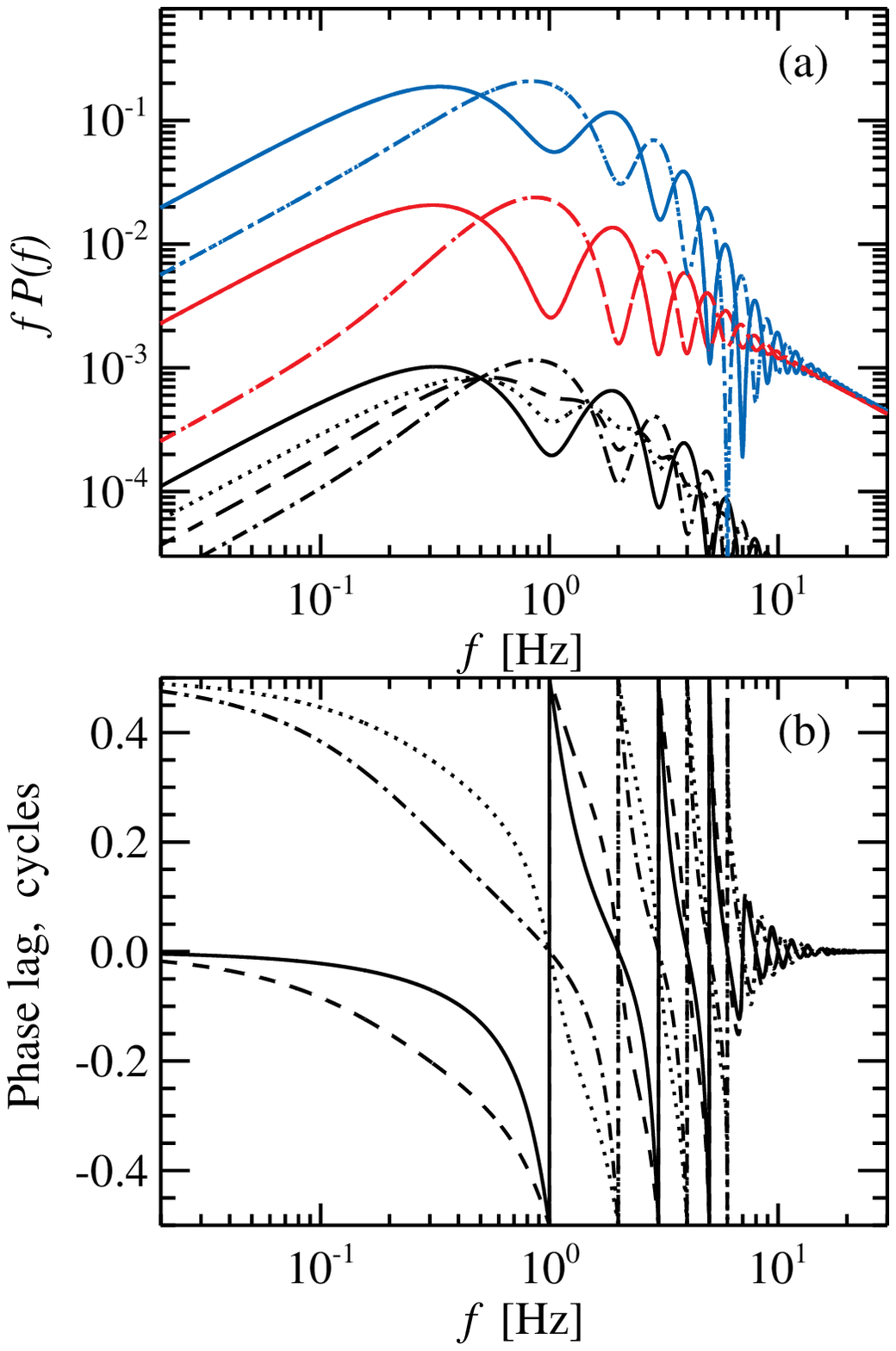}
\caption{
Same as in Fig.~\ref{fig:twocomp}, but for the inverse values of $ \varepsilon_{\rm s}$.
Parameters are listed in Table~\ref{tab:twocomp}.
}\label{fig:twocomp4} 
\end{figure}

\begin{table}
\caption{Parameters of models with two components in soft and hard X-rays (Figs~\ref{fig:twocomp}--\ref{fig:twocomp4}).
}\label{tab:twocomp}
  \begin{center}
\begin{tabular}{ccc}
\hline
\hline
  Line in Fig.~\ref{fig:twocomp}		&	$ \varepsilon_{\rm s}$	& $ \varepsilon_{\rm h}$	\\
\hline 
 solid			&	0.5		&	0.3					\\
 dotted		&	0.5	 	&      $-$0.3 		 \\
 dashed		&	$-$0.5	 	&      0.3   		 \\
 dot-dashed	&	$-$0.5	 	&      $-$0.3    	 \\
\hline
  Line in Fig.~\ref{fig:twocomp2}		&	$ \varepsilon_{\rm s}$	& $ \varepsilon_{\rm h}$	\\
\hline 
 solid			&	2		&	3.3					\\
 dotted		&	2	 	&      $-$3.3 		 \\
 dashed		&	$-$2	 	&      3.3   		 \\
 dot-dashed	&	$-$2	 	&      $-$3.3    	 \\
\hline
  Line in Fig.~\ref{fig:twocomp3}		&	$ \varepsilon_{\rm s}$	& $ \varepsilon_{\rm h}$	\\
\hline 
 solid			&	2		&	0.3					\\
 dotted		&	2	 	&      $-$0.3 		 \\
 dashed		&	$-$2	 	&      0.3   		 \\
 dot-dashed	&	$-$2	 	&      $-$0.3    	 \\
\hline
  Line in Fig.~\ref{fig:twocomp4}		&	$ \varepsilon_{\rm s}$	& $ \varepsilon_{\rm h}$	\\
\hline 
 solid			&	0.5		&	3.3					\\
 dotted		&	0.5	 	&      $-$3.3 		 \\
 dashed		&	$-$0.5	 	&      3.3   		 \\
 dot-dashed	&	$-$0.5	 	&      $-$3.3    	 \\
\hline
     \end{tabular}
  \end{center}
\end{table}

There are four alternatives for the signs of $\varepsilon_{\rm s}$ and $\varepsilon_{\rm h}$, each alternative is further divided into cases with ratios larger or smaller than unity and whether 
$\varepsilon_{\rm h}$ is larger or smaller than $\varepsilon_{\rm s}$.
As an illustration, we consider the case with single zero-centred Lorentzian $\Delta f = 0.5$~Hz, $t_0=0.5$~s, $f_{\rm filt}=5$~Hz and vary values of $\varepsilon_{\rm s}$ and 
$\varepsilon_{\rm h}$.
The resulting phase lags, power and cross-spectra are shown in Figs~\ref{fig:twocomp}--\ref{fig:twocomp4}.

In the frequency range without damping (where $G(f)=1$), the switch of $\varepsilon_{\rm s}$ and $\varepsilon_{\rm h}$ into $1/\varepsilon_{\rm s}$ and $1/\varepsilon_{\rm h}$, respectively, leads to the same power spectra, while the behaviour of the phase lags is more complex (compare the same lines in Figs~\ref{fig:twocomp} and \ref{fig:twocomp2}).
Namely, the phase lags change sign, $\Delta\varphi \to -\Delta\varphi$, if $\varepsilon_{\rm s}\varepsilon_{\rm h}>0$ (see solid and dot-dashed lines in Figs~\ref{fig:twocomp} and \ref{fig:twocomp2}), while for 
$\varepsilon_{\rm s}\varepsilon_{\rm h}<0$, the switch $\Delta\varphi \to \pi-\Delta\varphi$ occurs (dotted and dashed lines in Figs~\ref{fig:twocomp} and \ref{fig:twocomp2}).

As the damping factor $G(f)$ becomes more important, the role of the first (damped) term decreases, and the phase lags tend to zero at high frequencies.
There is a region, between 4$-$8~Hz in Fig.~\ref{fig:twocomp2}, where $\varepsilon_{\rm s}G(f)<1$, but $\varepsilon_{\rm h}G(f)>1$, for which the lags are monotonic functions of frequency (see below), however, as soon as the term $\varepsilon_{\rm h}G(f)$ becomes smaller than unity, the oscillations of phase lags become the same as those in Fig.~\ref{fig:twocomp}.
The power spectra in Fig.~\ref{fig:twocomp2} at high frequencies have substantially less power than those in Fig.~\ref{fig:twocomp}.
This is related to the relative contributions of the components, as for $\varepsilon_{\rm s}>1$ ($\varepsilon_{\rm h}>1$), the component giving dominant contribution is damped at high frequencies, while for $\varepsilon_{\rm s}<1$ ($\varepsilon_{\rm h}<1$) the damped component is only a small fraction of the undamped component.

If one of the ratios is larger than unity and the other is smaller than unity, the phase lags do not show oscillatory behaviour and instead demonstrate monotonic dependence on frequency (see Figs~\ref{fig:twocomp3} and \ref{fig:twocomp4}). 
Similar to the previous cases, the switch of $\varepsilon_{\rm h}\to1/\varepsilon_{\rm h}$ and $\varepsilon_{\rm s}\to1/\varepsilon_{\rm s}$ leads to a change $\Delta\varphi \to -\Delta\varphi$ for the case of $\varepsilon_{\rm s}\varepsilon_{\rm h}>0$ (solid and dot-dashed lines in Figs~\ref{fig:twocomp3} and \ref{fig:twocomp4}), while the switch $\Delta\varphi \to \pi-\Delta\varphi$ occurs for $\varepsilon_{\rm s}\varepsilon_{\rm h}<0$ (dotted and dashed lines in Figs~\ref{fig:twocomp3} and \ref{fig:twocomp4}).
The changes in $\Delta\varphi$ are caused by the changes of signs of real and imaginary part of the cross spectra (denominator and numerator of Eq.~\ref{eq:phase_lags_two}), which determine the quadrants the phase lags oscillate between.

We proceed to the cases when the soft and/or hard X-ray synchrotron Comptonization light curves are anti-correlated with the mass accretion rate (cases 2-4 in Fig.~\ref{fig:scheme}).
It can be shown that all these cases can be embedded into the developed formalism as follows.
The timing characteristics (PSDs, cross-spectra and phase lags) for the case 2, when only soft-band synchrotron Comptonization is anti-correlated with $\dot m(t)$, can be found by substituting $\varepsilon_{\rm s} \to - \varepsilon_{\rm s}$ into Eqs~(\ref{eq:cs_ampl_two}) and (\ref{eq:phase_lags_two}) and replacing the computed $\Delta\varphi \to \Delta\varphi \pm \pi$, where the choice of sign $\pm\pi$ in our formulation is determined in such a way that the phase lags at lowest frequencies belong to the interval $[-\pi;\pi)$.
For the case 3, when only hard-band synchrotron Comptonization light curve is anti-correlated with $\dot m(t)$, we need to replace $\varepsilon_{\rm h} \to - \varepsilon_{\rm h}$ and again use $\Delta\varphi \to \Delta\varphi \pm \pi$.
For the case 4, when both of synchrotron Comptonization light curves are anti-correlated with $\dot m(t)$, we put $\varepsilon_{\rm h} \to - \varepsilon_{\rm h}$ and $\varepsilon_{\rm s} \to - \varepsilon_{\rm s}$ to find the timing properties.

\section{Application to black hole binaries in the intermediate state}

The propagating fluctuations model quantitatively explains the power spectra and the phase lags in the soft and in the hard state of Cyg~X-1 \citep{RIvdKcygx1}, but experiences qualitative problems in predicting the observed characteristics during the intermediate state.
Moreover, the hard-state and intermediate-state data of XTE~J1550--564 are not well explained in the model of propagating fluctuations either \citep{RIvdK17j1550}.
We suggest the discrepancy is due to the presence of (at least) two distinct components in the Comptonization continuum.

The formalism developed in this work can be applied to the intermediate states of black hole X-ray binaries, when the X-ray spectrum can be composed of two components.
Both components carry the signatures of propagating fluctuations, however, due to the complex interconnection of the X-ray continua, the resulting power spectra and joint characteristics between the soft and hard bands deviate from the simple predictions of the propagating fluctuations model.
The cross-spectral amplitudes and phase lags can be obtained from Eqs.~(\ref{eq:cs_ampl_two}) and (\ref{eq:phase_lags_two}), and the power spectra -- from Eq.~(\ref{eq:psd_soft}).

\subsection{XTE~J1550$-$564}

% (Fig. 13)
\begin{figure}
\centering 
\includegraphics[width=7cm]{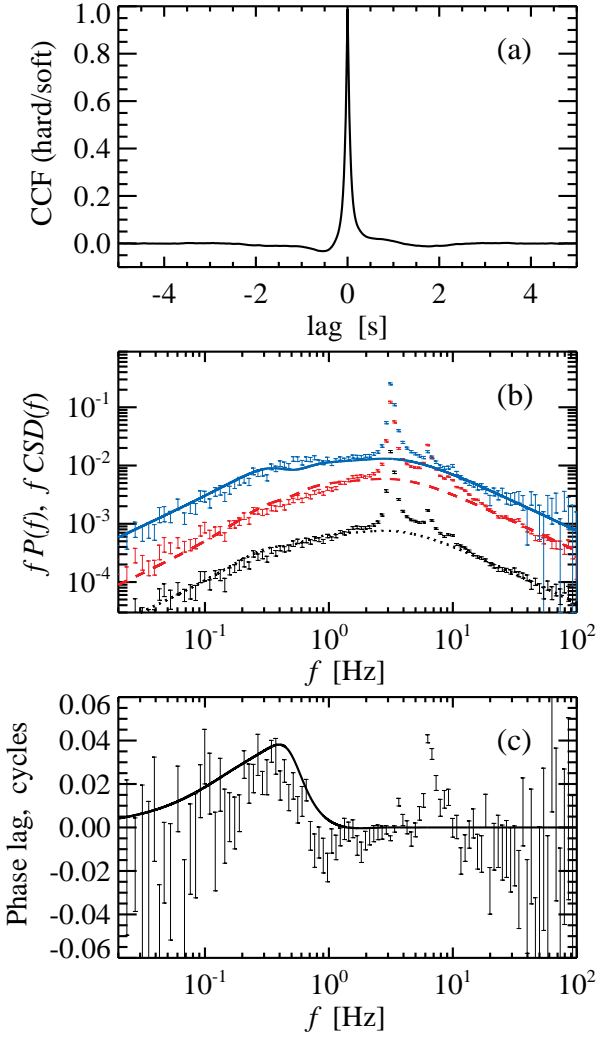}
\caption{
Example of model characteristics in application to the data on XTE~J1550$-$564.
The observed phase lags, power and cross-spectra are the same as shown in fig.~6 of \citet{RIvdK17j1550}, normalised according to \citet{MiKi89}.
(a) Cross-correlation function; (b) the hard-band (blue solid line and upper data points) and soft-band (red dashed line and data points in the middle) power spectra, and cross-spectrum (black dotted line and data points in the bottom), shifted down by a factor of 10 for clarity; (c) the phase lag spectrum. 
The model parameters are listed in Table~\ref{tab:par_xte1550cygx1}.
}\label{fig:xte1550} 
\end{figure}

% (Fig. 14)
\begin{figure}
\centering 
\includegraphics[width=7cm]{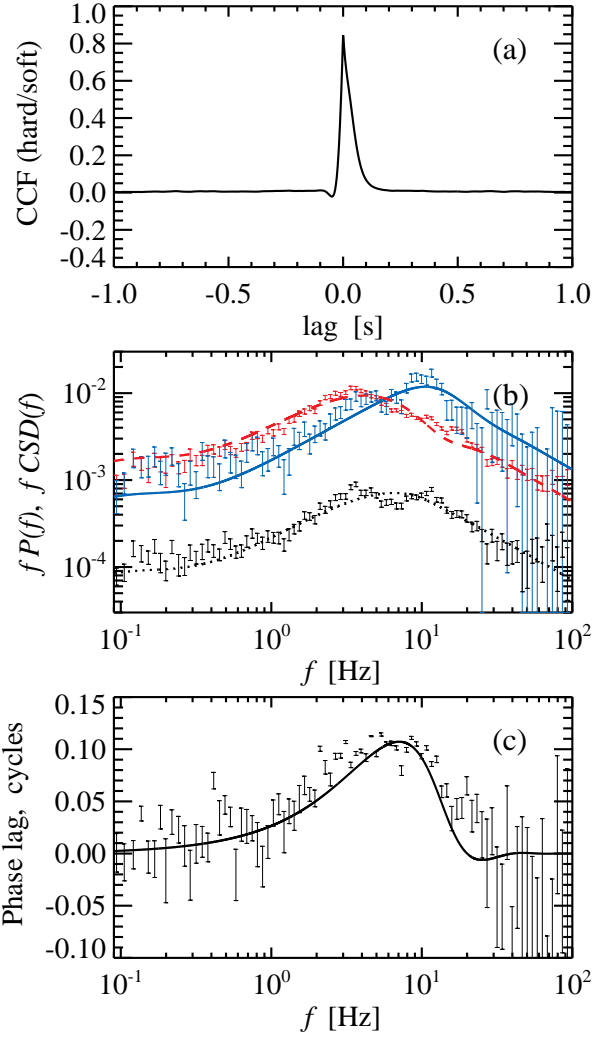}
\caption{
Same as in Fig.~\ref{fig:xte1550}, but in application to Cyg~X-1. 
Parameters are listed in Table~\ref{tab:par_xte1550cygx1} and data are the same as in fig.~4 of \citet{RIvdKcygx1}.
}\label{fig:cygx1} 
\end{figure}

Let us consider the specific case of the intermediate state of XTE~J1550$-$564 studied in \citet{RIvdK17j1550}, {\it RXTE} observation ID~30199-06-07-00.
The authors find that the soft band ($1.9-13$~keV) power spectrum lacks power at low frequencies, as compared to the hard band ($13.4-20.3$~keV), while the high-frequency part is the same for both bands.
In addition, the phase lags increase at low frequencies, reaching $\Delta \varphi /2\pi \approx 0.04$, and then decrease to zero for frequencies $f\geq1$~Hz.
Both bands may contain soft power-law (disc Comptonization) continuum, hard power-law (synchrotron Comptonization) continuum, as well as reflection from the cold disc.
The soft band may additionally contain contribution from the cold accretion disc itself.
We consider only possible contribution from the two Comptonization continua and use the developed formalism for the joint characteristics of the light curves in these bands.

We assume the power spectrum of the mass accretion rate fluctuations is a sum of two zero-centred Lorentzians with widths $\Delta f_1=0.5$~Hz and $\Delta f_2 = 3$~Hz and the squire of their rms ratio $r_2^2/r_1^2=4$. 
Though clearly an oversimplification to the power spectrum coming from propagating fluctuations, its shape mimics the power-law $\propto f^{-1}$ in the frequency range $0.5-5$~Hz, similar to the observed one.
The normalisation of the power spectra depend on the relative variability amplitude, and for the soft band it is expected to be smaller than for the hard band, due to the presence of the stable cold accretion disc component.
In the proposed model, the normalisation is a free parameter.
The resulting cross-correlation function, cross- and power spectra and phase lags are shown in Fig.~\ref{fig:xte1550} and the parameters are listed in Table~\ref{tab:par_xte1550cygx1}.
We note that due to oversimplification of the underlying power spectrum, we only show qualitative agreement between the model and the data and do not aim to achieve the quantitative fit.
We also do not include the quasi-periodic oscillation.

\begin{table}
\caption{Model parameters for Figs~\ref{fig:xte1550}~and~\ref{fig:cygx1}.
}\label{tab:par_xte1550cygx1}
  \begin{center}
\begin{tabular}{ccc}
\hline
\hline
{Parameter}				& {XTE~J1550--564}		& {Cyg~X-1}		\\
\hline 
$r_2^2/r_1^2$				&  4						& 8		\\
$\Delta f_1$~(Hz)			&  0.5					& 0.1		\\
$\Delta f_2$~(Hz)			&  3						& 5		\\
$ \varepsilon_{\rm s}$		& $-$1.7					& 0.35		\\
$ \varepsilon_{\rm h}$		& $-$2.2					& $-$0.45		\\
 $t_0$~(s)			 		& 0.35					& 0.025	\\
 $f_{\rm filt}$~(Hz)			& 0.5						& 13	\\
\hline
     \end{tabular}
  \end{center}
\end{table}

The reduction of power at low frequencies is due to the partial cancellation of the variability, because fluctuations of the two X-ray components come out of phase. 
Because the ratio of components in the soft band $ \varepsilon_{\rm s}$ is closer to (minus) unity than $ \varepsilon_{\rm h}$, the reduction of power is more pronounced in the soft band than in the hard band.
The filtering frequency can be found from the characteristic frequency at which the phase lags approach zero.
Because $f_{\rm filt}$ is low, there is practically no interference picture seen in the power and cross spectra.

The maximal amplitude of the phase lags depends on the ratio of synchrotron and disc Comptonization continua according to Eq.~(\ref{eq:deltaphimax_two}), and for the chosen 
$\varepsilon_{\rm s}$ and $ \varepsilon_{\rm h}$ reaches the required $\Delta \varphi = 0.04$ cycles.
The frequency at which the maximum is reached can be obtained from Eq.~(\ref{eq:freq_maxmin_two}) and mostly depends on the delay. 
The obtained delay $t_0=0.35$~s is similar to the characteristic viscous timescales found in other objects with multiple humps in the power spectra \citep{V16}.

We note that the assumption of two components in both energy bands is essential for the considered observation.
In order to have the suppression of power at lower frequencies, as observed in the soft band, one needs to have an anti-correlation of two components. 
In this case, the model of two continua only in the soft ban predicts negative phase lags at low frequencies (see Fig.~\ref{fig:anti}), contrary to the observed phase lags.
Hence, the simpler model cannot explain the observed characteristics.

\subsection{Cyg~X-1}

Another example of model application is the power spectra of Cyg~X-1, {\it RXTE} observation ID~10412-01-07-00.
Multiple humps in power spectra, appearing at different frequencies in soft and hard bands, as well as the large value of the phase lag (achieving 0.1 cycles) resulted in an unacceptable fit with the propagating fluctuations model \citet{RIvdKcygx1}.
In Fig.~\ref{fig:cygx1} we show the power spectra, cross-spectra (shifted down by a factor of 10 for clarity) and phase lags which are similar to the observed ones in this object, along with the expected CCF.
We list the parameters in Table~\ref{tab:par_xte1550cygx1}.

As compared to the case of XTE~J1550$-$564, the delay is substantially smaller, while the damping frequency $f_{\rm filt}$ is substantially higher.
These trends are consistent with the expectation of the truncated disc scenario: for higher truncation radius, the delay is expected to be longer, while the damping frequency is smaller.
With the decrease of truncation radius, one would expect the increase of the disc to synchrotron Comptonization ratio ($\varepsilon_{\rm s}$ and $\varepsilon_{\rm h}$), however, we find their values for Cyg~X-1 to be lower than for XTE~J1550--564.
This discrepancy might be explained by the difference in spectral formation in these two sources.
Alternatively, this might be due to the physical drop of the absolute variability amplitude of the disc Comptonization component (or, equivalently, the increase of synchrotron Comptonization variability amplitude) during the hard to soft state transition.
Detailed investigation of the evolution of timing characteristics during the transition is needed to trace the changes of parameters and to understand physical origin of these changes.

The model with two components only in the soft band experiences substantial problems explaining the observed timing characteristics for these observations.
In this model, the hard band is produced by one component, hence its power spectrum reflects that of propagating fluctuations.
The hard-band power peaks at frequencies $\sim$10~Hz, while the soft-band power has a peak at lower frequencies.
The enhanced power at low frequencies can be explained by the artificial removal of power at higher frequencies (due to interference), and the absence of significant bumps at higher frequencies 
can be explained by the relatively low filtering frequency ($\sim$4--5~Hz).
However, such low filtering frequency is inconsistent with presence of significant phase lags at frequencies up to $\sim$20~Hz.
Thus, we conclude that the simpler model is not capable of reproducing the observed characteristics.

\section{Summary}\label{sect:summ}

We propose an analytical model for joint timing characteristics of the light curves containing two spectral components.
Presence of two components in the light curves cause interference picture in their power and cross-spectra and oscillatory behaviour of the phase/time lags.
We deduce simple analytical formulae to account for the coupling terms between the components.
In this model, the maximum value of the phase lag depends on the relative contribution of the components, but not on the delay between them. 
However, the Fourier frequencies at which the peaks of the phase lags are reached, do depend on the delay.
The interference redistributes power between Fourier frequencies and, depending on the relative contribution of spectral components and their relative phase of variability, the power can be substantially reduced at low frequencies.
This may cause the deviation from the linear rms--flux relation observed in the intermediate state.

The results can be used to study variability properties in two or more X-ray energy bands in accreting black holes and neutron stars, which is particularly promising in the light of the future high-time resolution X-ray missions, such as {\it STROBE-X}, but also can be applied to the light curves observed at longer wavelengths (UV, optical, infrared).
Using the developed formalism, we reproduce the soft and hard X-ray band timing properties of two black hole X-ray binaries in the intermediate state, where the standard model of propagating fluctuations fails to predict the power spectral shape and large value of the phase lags between the bands.

We note that further complications for the power spectra and phase lags are expected in the case when the spectral changes cannot be described by simple spectral pivoting (changes of spectral index at constant normalisation) and/or changes in the overall normalisation, as well as if the variability amplitudes themselves depend on the mass accretion rate.
These cases can be studied in the spectral-timing approach, when the light curves are produced from the direct modelling of spectral formation in the medium with changing parameters, which we plan to study in future works.

\section*{Acknowledgements}

The work was supported by the Academy of Finland grant 309308 and the Ministry of Education and Science of the Russian Federation grant 14.W03.31.0021.
AV thanks Stefano Rapisarda for sharing the data, Juri Poutanen for the comments on the manuscript and the referee for the useful suggestions.

%\bibliographystyle{mnras}
%\bibliography{../allbib}
%\end{document}

\end{document}